\documentclass[12pt,English,
urlcolor=black,linkcolor=black]{article}

\usepackage[dvips]{graphicx}
\usepackage{epsfig}
\usepackage{amsmath,amsfonts,amssymb,amsthm}
\usepackage{mathrsfs,mathtools}
\usepackage{verbatim}
\usepackage{psfrag}
\usepackage{bm}
\usepackage{bbm}
\usepackage[utf8]{inputenc}
\usepackage[square,comma,sort&compress,numbers]{natbib}
\usepackage{color}
\usepackage{slashed}
\usepackage{upgreek}
\usepackage{enumitem}
\usepackage{float}
\usepackage{soul}

\usepackage[colorlinks,citecolor=black]{hyperref}

\usepackage{epsf,epsfig}
\usepackage{graphics}
\usepackage{subcaption}

\newcommand\blfootnote[1]{%
  \begingroup
  \renewcommand\thefootnote{}\footnote{#1}%
  \addtocounter{footnote}{-1}%
  \endgroup
}

\def\d{\mathrm{d}}

\def\i{\mathrm{i}}
\def\e{\mathrm{e}}

\usepackage[errorstop]{feynmp}

\begin{document}

\thispagestyle{empty}

\begin{flushright}
{
\small
CP3-20-25
}
\end{flushright}

\vspace{0.4cm}

\begin{center}
\Large\bf\boldmath
Nonsingular black hole in two-dimensional asymptotically flat spacetime
\unboldmath
\end{center}

\vspace{0.2cm}

\begin{center}
{Wen-Yuan Ai$^*$\blfootnote{$^*$wenyuan.ai@uclouvain.be}

\vskip0.2cm
{\it Centre for Cosmology, Particle Physics and Phenomenology,\\ Université catholique de Louvain, Louvain-la-Neuve B-1348, Belgium}\\

\vskip1.4cm}

\end{center}

\begin{abstract}

We study a special two-dimensional dilaton gravity with Lagrangian $\mathcal{L}=\frac{1}{2}\sqrt{-g}(\phi R+{\lambda^2}{\rm sech}^2\phi)$ where $\lambda$ is a parameter of dimension mass. This theory describes two-dimensional spacetimes that are asymptotically flat. Very interestingly, it has an exact solution for the metric,  $\d s^2=-(c+\tanh \lambda x)\d t^2+ \d x^2/(c+\tanh {\lambda} x)$, parametrized by $c$. For $c\in(-1,1)$, the solution presents an event horizon but no singularity. Because of the kink profile for the metric components appearing in the solution, we refer to the spacetime described by the metric as a {\it gravitational domain wall} with the wall simply being the event horizon and separating two asymptotically Minkowskian spacetimes. The global causal structure for such an object is studied via coordinate extension and the thermodynamical quantities are computed. Only when $c\in(-1,0]$ are the entropy and energy non-negative. 

\end{abstract}

\newpage
%\pacs{05.30.-d, 11.30.Fs, 12.60.Fr, 14.60.St, 95.30.Cq, 98.80.-k}
\tableofcontents

\section{Introduction}
\label{sec:intro}
Two-dimensional (2D) gravity theories are interesting because they allow one to isolate some important features of the higher-dimensional gravity while being very simple. They are therefore viewed as a theoretical laboratory for quantum gravity. In recent years, motivated by the Sachdev-Ye-Kitaev model~\cite{Sachdev:1992fk,K,Kitaev:2017awl} (see also Refs.~\cite{Maldacena:2016hyu,Sarosi:2017ykf,Trunin:2020vwy}), Jackiw-Teitelboim (JT) gravity~\cite{Jackiw:1984je,Teitelboim:1983ux,Almheiri:2014cka} has received a resurgence of interest, see, e.g., Refs.~\cite{Jensen:2016pah,Maldacena:2016upp,Engelsoy:2016xyb,Harlow:2018tqv,Saad:2019lba,Stanford:2019vob,Iliesiu:2020zld}. JT gravity belongs to the family of two-dimensional  dilaton gravities that can be parametrized by the following general action\footnote{In general, the $\phi$ in the first term in the square brackets can be a general function $Z(\phi)$. But for physical reasons, $Z'(\phi)$ is nonzero everywhere and we can introduce a new field to put the action into the form of Eq.~\eqref{eq:dilaton-gravities}. Moreover, here we only consider torsionless gravity. For reviews of dilaton gravity, see Refs.~\cite{Nojiri:2000ja,Strobl:1999wv,Grumiller:2002nm,Grumiller:2006rc}, and for recent studies, see Refs.~\cite{Bhattacharya:2020qil,Narayan:2020pyj}.}
\begin{align}
\label{eq:dilaton-gravities}
S=\frac{1}{2}\int\d^2 x\sqrt{-g}\left[\phi R-U(\phi)g^{\alpha\beta}(\partial_\alpha\phi)\partial_\beta\phi+W(\phi)\right],
\end{align}
where $g_{\alpha\beta}$ is the 2D spacetime metric. JT gravity corresponds to $U(\phi)=0$, $W(\phi)=2\Lambda\phi$. In this case, $\phi$ is an auxiliary field and can be eliminated by its algebraic equation of motion, giving $R=-2\Lambda$. Depending on the sign of $\Lambda$, JT gravity thus describes a 2D de Sitter or anti-de Sitter space. For convenience, one usually takes $|\Lambda|=1$ so that $\Lambda=\pm 1$.{\footnote{Since $\Lambda$ is a dimensionful parameter, one shall restore it in physical quantities via the dimensional analysis.}} Equation~\eqref{eq:dilaton-gravities} also contains other interesting models as its special cases. For instance, the Callan-Giddings-Harvery-Strominger model~\cite{Callan:1992rs} corresponds to $U(\phi)=-\frac{1}{\phi}$, $W(\phi)=-4\lambda^2\phi$ and the spherically reduced gravity~\cite{Berger:1972pg,Benguria:1976in,Thomi:1984na,Hajicek:1984mz} is given by 
\begin{subequations}
\begin{align}
U(\phi)&=-\frac{(D-3)}{(D-2)\phi},\\ W(\phi)&=-\lambda^2(D-2)(D-3) \phi^{\frac{(D-4)}{(D-2)}},
\end{align}
\end{subequations}
where $D>2$ is an integer representing the dimension of the spacetime before the spherical reduction. In both cases $\lambda$ is a parameter of dimension one in mass. 

Actually, with a Weyl transformation, the factor $U(\phi)$ in Eq.~\eqref{eq:dilaton-gravities} can always be set to zero and hence 2D dilaton gravity can be generally described by~\cite{Banks:1990mk}\footnote{Note that when $U(\phi)$ is singular the two theories before and after the transformation are not necessarily equivalent~\cite{Grumiller:2002nm}.}
\begin{align}
\label{eq:dilaton-simplified}
S=\frac{1}{2}\int\d^2 x\sqrt{-g}\left[\phi R+W(\phi)\right].
\end{align}
Varying the action with respect to $\phi$, one has
\begin{align}
R=-\frac{\d W(\phi)}{\d\phi}.
\end{align}
In this paper, we consider a special model with 
\begin{align}
\label{eq:model}
W(\phi)={\lambda^2}{\rm sech}^2\phi,
\end{align}
{where $\lambda$ is a parameter of dimension mass. As done for JT gravity, we will also set $\lambda=1$ and one should be able to restore it in physical quantities.} 
As we shall see shortly, this theory describes asymptotically flat spacetimes. We will show that it permits a very interesting solution with a free parameter in which there is an event horizon but no singularity. The structure is similar to the domain wall of a scalar field theory. Therefore we refer to this new gravitational object as a {\it gravitational domain wall} (or gravitational kink) with the ``wall'' representing the event horizon. We will then study its causal structure and thermodynamical properties. We shall see that for a particular choice of the free parameter in the solution, the gravitational domain wall can have a nonzero temperature while its energy and entropy are vanishing. Unlike JT gravity, which is related to the near-horizon region of extremal black holes in higher-dimensional spacetime through spherical reduction, it is not clear how the potential~\eqref{eq:model} arises from a higher-dimensional theory. Nonetheless, given the interesting properties mentioned above and its potential role in building holography for asymtotically flat two-dimensional spacetimes, it still deserves to be investigated.  

The remainder of this article is organized as follows. In the next section we study the 2D dilaton gravity with the potential~\eqref{eq:model} and show that it has exact solutions in which the zero-zero metric component takes the kink profile when using the Schwarzschild-like ansatz. We then study the global causal structure for the obtained solutions via coordinate extension. In Sec.~\ref{sec:thermodynamic}, the thermodynamical properties for the gravitational domain wall are investigated. We conclude in Sec.~\ref{sec:conclusions}.

\section{Gravitational domain wall in two-dimensional dilaton gravity}\label{sec:domain-wall}

\subsection{Equations of motion and solutions}

A general classical solution in the theory given by Eq.~\eqref{eq:dilaton-simplified} can be put in the form~\cite{Gegenberg:1994pv,Witten:2020ert}
\begin{align}
\label{eq:Schwarzschild-ansatz}
\d s^2=-A(x)\d t^2+\frac{1}{A(x)}\d x^2.
\end{align}
Note that whether or not one shall interpret $t$ as time and $x$ as space depends on the sign of $A(x)$. If $A(x)>0$ then Eq.~\eqref{eq:Schwarzschild-ansatz} describes a static spacetime region while if $A(x)<0$ it describes a time-dependent spatially homogeneous region. Following Ref.~\cite{Witten:2020ert}, one can obtain one of the dynamic equations by first considering the metric 
\begin{align}
\label{eq:AG}
\d s^2=-A(x)\d t^2+\frac{1}{G(x)}\d x^2,
\end{align} 
and then substituting $G(x)=A(x)$ into the equation of motion for $G(x)$. From Eq.~\eqref{eq:AG}, one has the Ricci scalar
\begin{align}
\label{eq:R}
R=-\frac{A'G'}{2A}-\frac{GA''}{A}+\frac{GA'^2}{2A^2},
\end{align}
where a prime denotes the derivative with respect to $x$. Substituting Eq.~\eqref{eq:R} into the action~\eqref{eq:dilaton-simplified}, one obtains
\begin{align}
\label{eq:actionGA}
S=\frac{1}{2}\int\d t\d x \left(-\phi\frac{\d}{\d x}\left(\sqrt{\frac{G}{A}}\frac{\d A}{\d x}\right)+\sqrt{\frac{A}{G}}W(\phi)\right),
\end{align}
from which the algebraic equation of motion for $G(x)$ reads
\begin{align}
\frac{\phi'A'}{\sqrt{AG}}-\frac{\sqrt{A}}{G^{3/2}}W(\phi)=0.
\end{align}
Now substituting $G=A$ into the above equation, one finally arrives at 
\begin{align}
\label{eq:eom1}
\phi'A'-W(\phi)=0.
\end{align}
To obtain the other equation of motion, we let $G=A$ in the action~\eqref{eq:actionGA} and get
\begin{align}
S=\frac{1}{2}\int\d t\d x \left(-\phi A''+W(\phi)\right).
\end{align}
Varying the action with respect to $A$, we have
\begin{align}
\label{eq:eom2}
\phi''=0.
\end{align}

From Eq.~\eqref{eq:eom2}, one has $\phi=ax+b$ where $a$, $b$ are constants. If $a=0$, then from Eq.~\eqref{eq:eom1} we have $W(\phi)=0$. For the model we consider [cf. Eq.~\eqref{eq:model}], $W(\phi)$ vanishes only at infinity, $\phi=\pm\infty$. And in this case, $R=-\d W(\phi)/\d \phi|_{\phi=\pm\infty}\equiv 0$. Therefore, this solution describes the trivial Minkowski spacetime, $A(x)={\rm constant}\neq 0$. 

If $a\neq 0$, then one can always let 
\begin{align}
\phi=x
\end{align}
after a transformation $(t,x,A)\rightarrow (t/\tilde{a}, \tilde{a}x+\tilde{b},\tilde{a}^2A)$ ($\tilde{a}\neq 0$), which leaves the metric~\eqref{eq:Schwarzschild-ansatz} invariant. In this case, Eq.~\eqref{eq:eom1} reduces to 
\begin{align}
\frac{\d A(x)}{\d x}={\rm sech}^2x,
\end{align}
from which one obtains $A(x)=(\tanh x)+c$ with $c$ being a constant. Substituting $A(x)=(\tanh x)+c$ and $A(x)=G(x)$ into Eq.~\eqref{eq:R}, one confirms that $R=2({\rm sech}^2x) \tanh x=-W'(x)$, which vanishes at $|x|\rightarrow\infty$. Thus the above solutions parametrized by $c$ describe asymptotically flat spacetimes. The metric reads
\begin{align}
\label{eq:gravdomainwall}
\d s^2=-(c+\tanh x)\d t^2+\frac{1}{c+\tanh x}\d x^2.
\end{align}
If $c>1$ or $c<-1$, there is no solution for $A(x)=0$ and thus there is no horizon for the spacetime considered. For $c\in (-1,1)$, the horizon is located at a finite distance
\begin{align}
x_h=-{\rm artanh}\, c,
\end{align}
while for $c=-1$ or $c=1$, the horizon is located at $+\infty$ or $-\infty$, respectively. Below we shall consider the case $c\in (-1,1)$.

\begin{figure}[H]
\centering
\includegraphics[scale=.6]{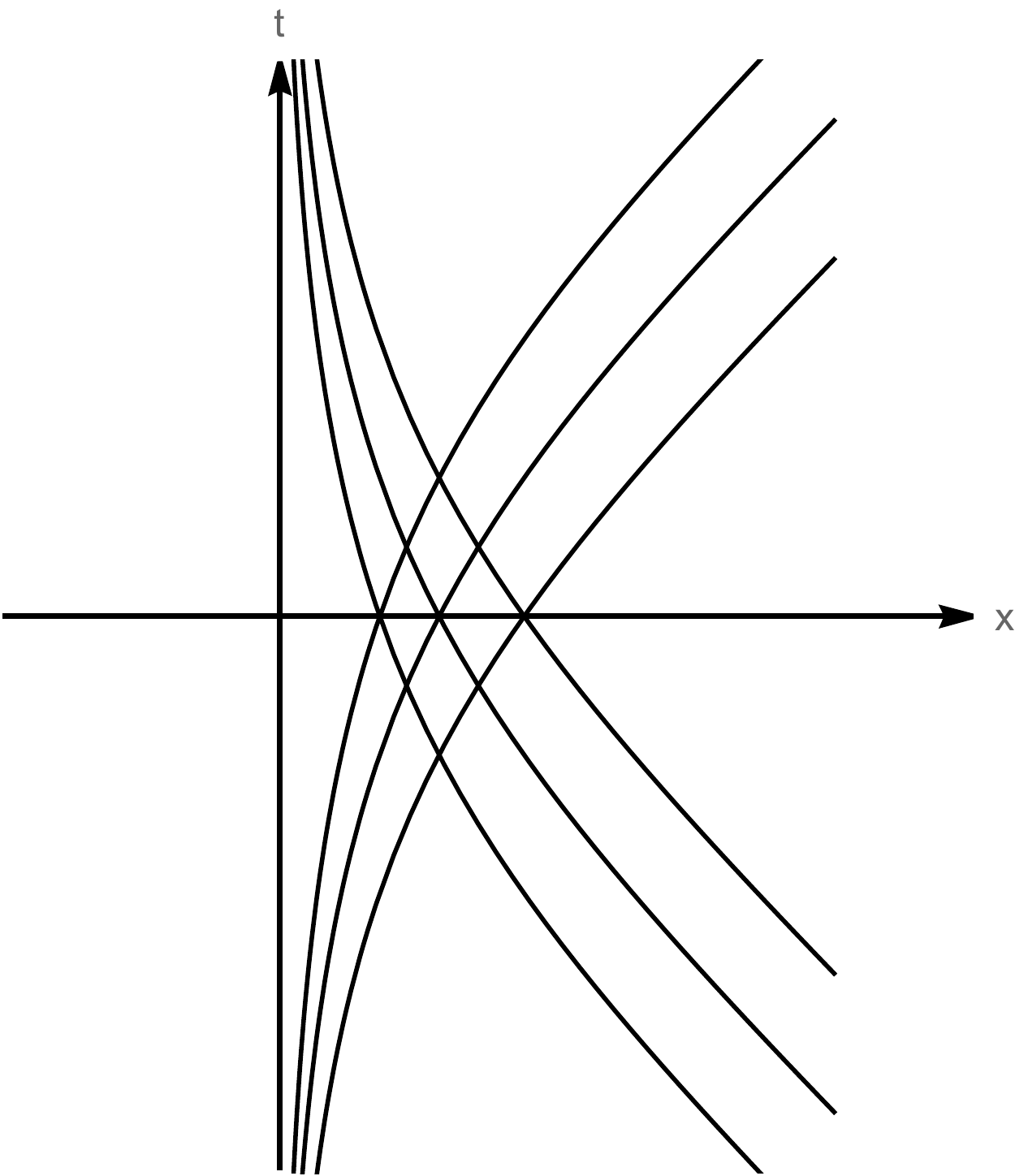}
\caption{The null geodesics outside of the gravitational domain wall for $c=0$.}
\label{fig:null}
\end{figure}

The coordinate singularity at $x=x_h$ is an event horizon. The region $x>x_h$ corresponds to the static region outside of the horizon while $x<x_h$ the region inside of the horizon which is, however, time dependent. Note that one shall interpret $x$ as the time for $x<x_h$. The metric has no essential singularity so that it describes a novel gravitational structure where there is an event horizon but no geometrical singularity. Because of the kink solution, $c+\tanh x$ in Eq.~\eqref{eq:gravdomainwall}, we refer to this object as gravitational domain wall---the wall is simply the event horizon. The null geodesics for $x>x_h$ are given by
\begin{subequations}
\begin{align}
&t=\frac{c x-\log(c\cosh x+\sinh x)}{-1+c^2}+s,\quad {\rm for\ outgoing\  light\ beams,}\\
&t=-\frac{c x-\log(c\cosh x+\sinh x)}{-1+c^2}+s,\quad {\rm for\ ingoing\  light\ beams},
\end{align}
\end{subequations}
where $s$ is a constant. We plot the null geodesics outside of the horizon for $c=0$ in Fig.~\ref{fig:null}. For the general case, the null geodesics look similar but with the horizon horizontally shifted from $x=0$.

\subsection{Coordinate extension}

To see that the wall is indeed an {\it event horizon}, we now do the coordinate extension starting from $x>x_h$. We first introduce the tortoise coordinate for our spacetime
\begin{align}
x_*=\frac{c x-\log(c\cosh x+\sinh x)}{-1+c^2}
\end{align}
so that the metric can be written as
\begin{align}
\d s^2=-(c+\tanh x)\left(\d t^2-\d x_*^2\right).
\end{align}
Note that $x_*\in (-\infty,\infty)$. Further, we introduce the retarded and advanced coordinates $\{u,v\}$ as
\begin{align}
u=t-x_*,\ v=t+x_*,\ -\infty<u, v<\infty.
\end{align}
Based on them we introduce 
\begin{align}
U=-\e^{-u/2},\ V=\e^{v/2}.
\end{align}
For now, $U\in(-\infty,0)$, $V\in(0,\infty)$. In terms of the coordinates $\{U,V\}$, the metric can be written as
\begin{align}
\d s^2=-\frac{4\e^{-cx}}{(1-c^2)^2\cosh x}\d V\d U.
\end{align}
Now we can extend the range for $U,V$ to $\mathbb{R}$. One can also write the above metric in a form
\begin{align}
\label{eq:kruskal}
\d s^2=\frac{4\e^{-cx}}{(1-c^2)^2\cosh x}\left(-\d T^2+\d X^2\right)
\end{align}
using the coordinates $T=\frac{1}{2}(V+U)$, $X=\frac{1}{2}(V-U)$. From the coordinate transformations, one has the relation
\begin{align}
\label{eq:relationXTx}
X^2-T^2=\e^{-cx}\left(c\cosh x+\sinh x\right).
\end{align}

\begin{figure}[H]
\centering
\includegraphics[scale=0.32]{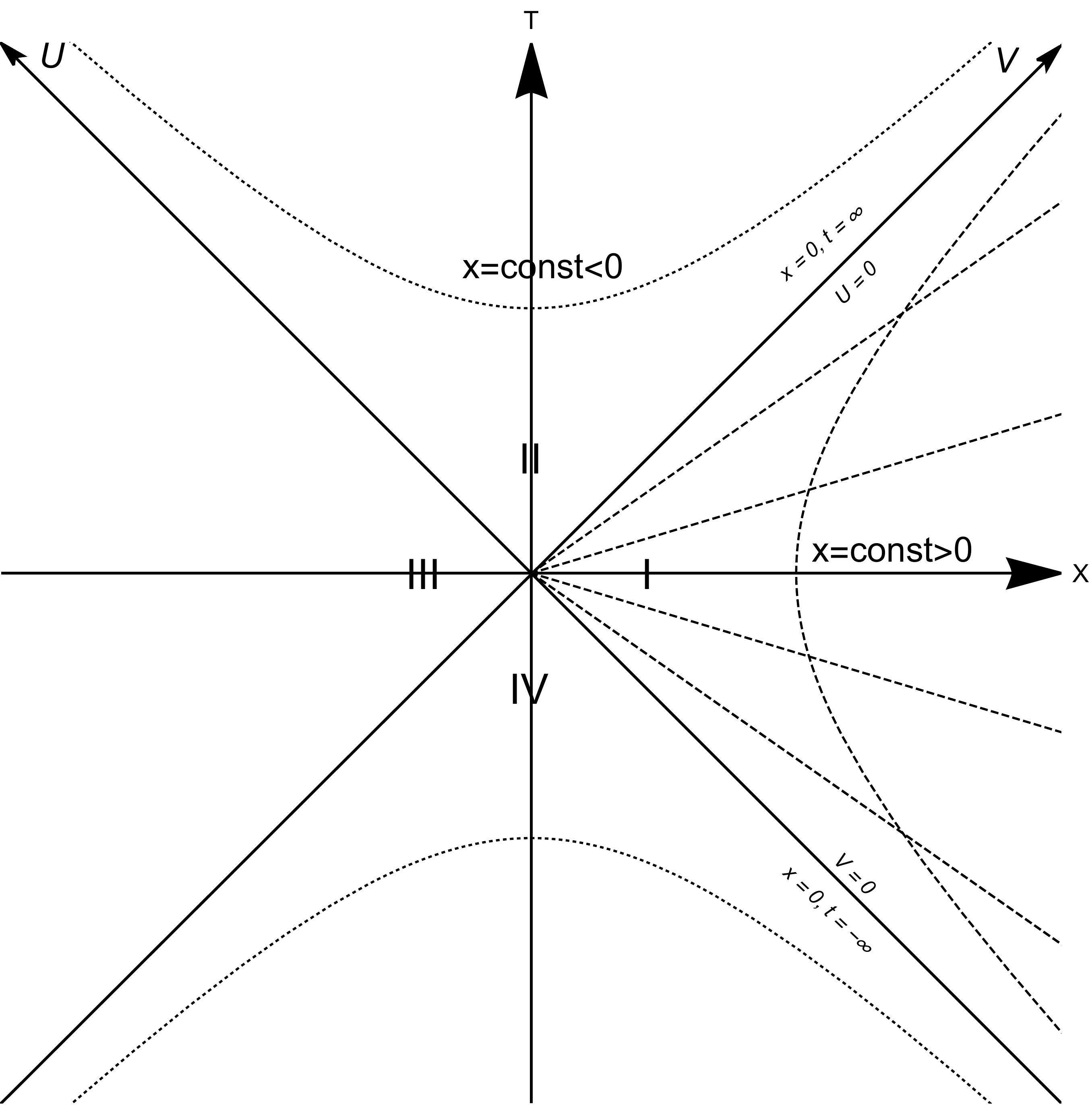}
\caption{The spacetime extended from the $x>x_h$ region in Eq.~\eqref{eq:gravdomainwall}. The hyperbolic curves are given by constant $x$ with $x>x_h$ outside of the horizon and $x<x_h$ inside of the horizon. We identify regions I and II as, respectively, the regions $x>x_h$ and $x<x_h$ in our original spacetime~\eqref{eq:gravdomainwall}.}
\label{fig:kruskal}
\end{figure}

\begin{figure}[H]
\centering
\includegraphics[scale=.7]{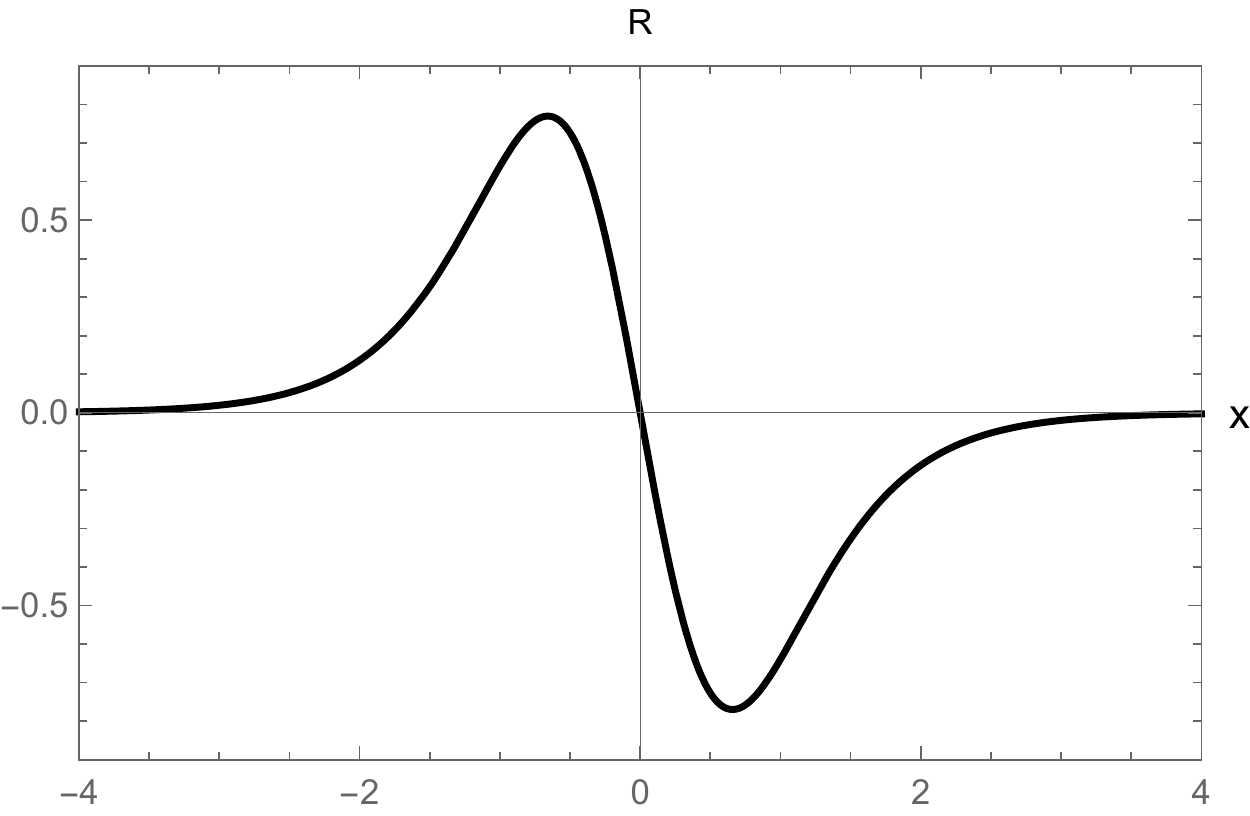}
\caption{The Ricci scalar for the gravitational domain wall as a function of $x$ for $c=0$. Note that only for $x>0$, $x$ can be interpreted as the spatial coordinate.}
\label{fig:Ricci}
\end{figure}

We show the extended spacetime in Fig.~\ref{fig:kruskal}. Region I corresponds to $x>x_h$ where the extension starts while region II represents $x<x_h$ in our original spacetime. Since in the coordinate system $\{T,X\}$ the null geodesics are always at $\pm 45^{\rm o}$ to the vertical, it is clear that no signal can escape from region II to region I. Hence, $x=x_h$ is indeed an event horizon. The spacetime~\eqref{eq:gravdomainwall} thus can be viewed as a black hole {\it without singularity}.\footnote{It seems that one  can instead identify the region $x<0$ as the ``white hole'' region (region IV in Fig.~\ref{fig:kruskal}) in the extended spacetime. We choose the other identification as it may shed new insights on black holes and Hawking radiation.} Our spacetime is certainly different from Rindler spacetime because the Ricci scalar is not trivially vanishing everywhere. In Fig.~\ref{fig:Ricci}, we plot the Ricci scalar as a function of $x$ for $c=0$. The Ricci scalar vanishes at $|x|=\pm\infty$ and the horizon.

\section{Thermodynamical properties of the gravitational domain wall}
\label{sec:thermodynamic}

In this section, we will study the thermodynamical properties of the gravitational domain wall through the Euclidean method~\cite{Gibbons:1976ue,Hawking:1982dh,Witten:2020ert}. The Euclidean action is
\begin{align}
\label{eq:Eucaction}
I_b=-\frac{1}{2}\int\d^2 x\sqrt{g}\left[\phi R+W(\phi)\right],
\end{align} 
where $x^\mu=\{\tau,x\}$ and all the quantities are Euclidean. The equations of motion remain to be the same as those in Minkowski space [Eqs.~\eqref{eq:eom1} and~\eqref{eq:eom2}]. The only difference is that in Euclidean space we now have $x\geq x_h$. [This can be seen if we perform a Wick rotation on the coordinate $T$ with $T\rightarrow -\i\mathcal{T}$ and we find from Eq.~\eqref{eq:relationXTx} that $x\geq x_h$.] From now on we label $x-x_h$ as $r$. Thus we have $\phi=r+x_h$, $A(r)=c+\tanh (x_h+r)$. Near the horizon, 
\begin{align}
A(r)=({\rm sech}^2x_h) r+\mathcal{O}(r^2).
\end{align}
Defining $y=\sqrt{r}$, we have the metric
\begin{align}
\d s^2=\frac{4}{{\rm sech}^2x_h}\left(\d y^2+\frac{({\rm sech^4}x_h)y^2}{4}\d\tau^2\right).
\end{align}
Only if $\tau$ has a period $\beta=4\pi/{\rm sech}^2 x_h$, the conical singularity at $y=0$ disappears. Therefore, we find
\begin{align}
\label{eq:temp}
T_{\rm H}=\frac{1}{\beta}=\frac{{\rm sech}^2 x_h}{4\pi}=\frac{{\rm sech}^2 (-{\rm artanh}\,c)}{4\pi}.
\end{align} 
The locally measured temperature is $T_{\rm loc}=T_{\rm H}/(\sqrt{c+\tanh x})$.

To derive the energy and the entropy, we use $Z=\exp(-\beta F)$ where $Z$ is the Euclidean partition function and $F\equiv E-TS$ is the free energy (here $E$ and $S$ denote the energy and the entropy, respectively). Using the semiclassical approximation for the partition function, $Z\approx \exp(-I)$ where $I$ is the total action for the classical background, one therefore arrives at 
\begin{align}
\label{eq:IES}
I=\beta E-S.
\end{align}
In the action $I$, the bulk action $I_b$ should be supplemented by the Gibbons-Hawking-York boundary term~\cite{York:1972sj,Gibbons:1976ue} which reads\footnote{For the asymptotically anti-de Sitter spaces, $C=-1$.}
\begin{align}
I_{\rm GHY}=-\int_{r=r_\infty}\d \tau\,\sqrt{h}\phi (K+C)
\end{align}
where $h=A(r_{\infty})=c+1$ is the induced metric at the boundary and $K$ is the extrinsic curvature of the boundary. Here $C$ is a term that depends only on the induced metric $h$ on the boundary but is independent of the metric $g$. In the case of asymptotically flat metrics, it is natural to choose $C$ so that the total action for the flat-space metric $\eta$ vanishes. Then 
\begin{align}
I_{\rm GHY}=-\int_{r=r_\infty}\d \tau\,\sqrt{h}\phi [K],
\end{align} 
where $[K]$ is the difference of the extrinsic curvature of the boundary in the metric $g$ and the metric $\eta$. For the flat-space metric, $K={1/r_\infty}$, while for the gravitational domain wall, $K=A'(r_{\infty})/(2\sqrt{A(r_\infty)})$. We thus obtain $I_{\rm GHY}=\beta\left(\sqrt{1+c}\right)$. Substituting the solution $A(r)$ into the bulk action one obtains\footnote{Note that there could be ambiguities in identifying $E$ and $S$ through Eq.~\eqref{eq:IES} since $\beta$ could have been substituted for its value in the intermediate steps. To get rid of such ambiguities, one should follow the procedure where $W(\phi)$, which determines the temperature, is taken as a general function. See, e.g., Ref.~\cite{Witten:2020ert}.} 
\begin{align}
I_b=-\frac{\beta}{2}\int_{x_h}^{x_\infty}\d x\, (-x A''(x)+{\rm sech}^2x)=-\beta(1+c)-2\pi x_h.
\end{align}
Substituting the obtained $I=I_{\rm GHY}+I_b$ into Eq.~\eqref{eq:IES} and comparing both sides, one obtains
\begin{align}
E=\sqrt{1+c}-(1+c),\quad S=2\pi x_h=2\pi\, {\rm artanh} (-c).
\end{align}
To have a non-negative entropy, we require $c\in(-1,0]$. In this physical parameter region, the energy is also non-negative. Note that when $c=0$, we have $E=S=0$ but we still have a horizon with a nonvanishing temperature.

Before ending this section, we make some comments below. The Hawking temperature has a general expression~\cite{Grumiller:2006rc,Bagchi:2014ava}
\begin{align}
T_{\rm H}=\frac{1}{2\pi}\left|\omega'(\phi)\right|_{\phi=\phi_h},
\end{align}
where 
\begin{align}
\omega(\phi)=-\frac{1}{2}\int^{\phi}\d y\, W(y)
\end{align}
and $\phi_h$ is the value for the dilation field at the horizon.
Substituting $W(y)={\rm sech}^2y$ into the above expressions, we obtain the same result as in Eq.~\eqref{eq:temp}. One could also introduce a more general model with the potential $W(\phi)={\rm sech^2}(\phi-\phi_0)$. Then the horizon would be located at $x=\phi=\phi_0-{\rm artanh} c$.

\section{Outlook and discussions}
\label{sec:conclusions}

In this paper, we propose a new 2D dilaton gravity whose action is given by Eq.~\eqref{eq:dilaton-simplified} with $W(\phi)={\lambda^2}{\rm sech^2\phi}$. This theory contains the 2D Minkowski space as its trivial solution. We report the discovery of a new gravitational structure in which there is an event horizon but no singularity. It is observed as a nontrivial solution in the theory we consider which reads
\begin{align}
\d s^2=-(c+\tanh {\lambda}x)\d t^2+\frac{1}{c+\tanh {\lambda}x}\d x^2,
\end{align}
where $c$ is a parameter. To have a physical horizon, $c\in(-1,0]$.
This metric has a coordinate singularity at $x=x_h\equiv {\rm artanh}(-c)$ which can be identified as an event horizon but contains no essential singularity.
Because of the kink profile in the metric, one may therefore refer to such an object as gravitational domain wall with the wall staying at $x_h$. We have studied the causal structure of the gravitational domain wall via coordinate extension and it is confirmed that no signal can escape from the left side of the wall to the right side. Therefore the left side of the wall can be viewed as the interior of a black hole and the right side as the exterior. We have also computed the thermodynamical quantities for the gravitational domain wall. It remains to be investigated whether or not this novel gravitational structure can shed some insights on Hawking radiation and the information paradox. 

{Some issues remain to be addressed. In particular, demonstrating that the theory considered here could arise from a higher-dimensional theory would largely improve the physical relevance of the gravitational domain wall. The case $c=0$ is of particular interest because it gives a horizon possessing a nonvanishing temperature but vanishing entropy and energy. It appears to be a bit strange and remains to be understood physically. Further, considering the simplicity of the model, it would be beneficial to consider the coupling to matter fields. As shown in Ref.~\cite{Grumiller:2006ja}, one can integrate out the metric and dilaton fields and obtain a nonlocal effective theory for the matter fields that can be reinterpreted in terms of the exchange of virtual black hole geometries. In view of our model having nonsingular black hole
solutions, the properties of the virtual black hole states could be particularly simple. It would be also interesting to work out the spectrum of the quasinormal modes.  We leave these issues for future work. }

%\acknowledgments

\appendix

\end{document}